\documentclass{desyproc}

\def\cascade{{\sc Cascade}}
\def\pythia{{\sc Pythia}}

\def\powheg{{\sc Powheg}}

\begin{document}


\title{Energy flow observables in hadronic collisions}



%
%
%
%
\author{{\slshape F.~Hautmann}\\[1ex]
  Department  of  Theoretical Physics, University of Oxford, Oxford OX1 3NP }
%
%
%
%
%



\contribID{ZZ}
\confID{UU}
\desyproc{DESY-PROC-2012-YY}
\acronym{MPI@LHC 2011}
\maketitle


\begin{abstract}
We   present  recent  QCD calculations of energy flow distributions 
associated with the production of jets at wide rapidity separations in high-energy 
hadron collisions,  and   discuss the 
role of these observables to analyze  contributions from parton showering and 
from multiple parton collisions.  
\vskip 0.8cm 
Contribution at  {\em Multiple Parton Interactions at the LHC 2011},  
Hamburg, November 2011  
\vskip -8.5cm 
\hspace*{10.0 cm} {  OUTP-12-09P }
\vskip 8.4cm 
\end{abstract}



\vskip 0.8 cm

Jet rates and event shape variables  have long been used~\cite{bryan-ts} 
to characterize  QCD  final states from    hard scatter events   
at  high-energy colliders 
and to describe 
    the event's energy flow.   Jet shape variables  describing  
     the jet's  internal structure and the  energy flow  within a  jet 
   have  also  been  studied,  
and  are being   proposed~\cite{jetsubs}  as diagnostic tools  at the LHC  in 
searches for  potential new physics signals      
from highly boosted massive states.  
In the last year first  LHC  measurements     
of     event shapes~\cite{evshape-cms}
         and jet shapes~\cite{jetshape-lhc}  have been performed.

In all these cases,  the interpretation of results depends on a good understanding of the 
overall structure of the final states.  This in turn 
implies  controlling  effects  due to strong interaction dynamics 
in the initial state.  Thus for instance jet shape  observables 
such as~\cite{substr-var,subjetty}  that are sensitive to the  jet's   substructure are also 
sensitive to soft physics effects, including the underlying event, pile-up,  and 
  multiple parton interactions~\cite{ajaltouni,bartal}.   
Hadronic event shapes measured at the LHC~\cite{evshape-cms}    
suggest    that   parton showering  effects  dominate 
   contributions of hard matrix elements   evaluated at high multiplicity. 

\begin{figure}[htb]
\vspace{25mm}
\includegraphics{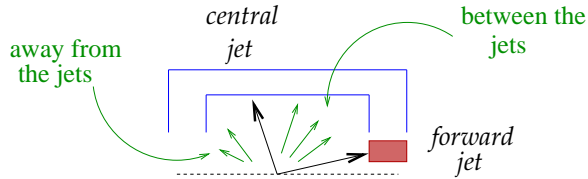}  
\caption{\it Production of forward and central jets: energy flow  
 in the inter-jet and outside 
regions. } 
\label{fig:jetcorr} 
\end{figure}

In this article we focus on parton showering and multi-parton interactions  
(for recent discussions reviewing  these topics, see respectively~\cite{hoeche11}    
and~\cite{markus-talk,yuri-talk}),   
and we  discuss   energy flow  observables~\cite{epjc12}  which  become 
 measurable,  essentially for the first time, at the LHC,   and may be used for studies 
  of showering and  of  multiple collisions.  
  The main focus is on     the region   of  high  rapidities,  
   where   production of final states 
   with sizeable momentum transfers   presents  new features  at  the 
   LHC compared to previous collider experiments~\cite{ajaltouni}.    
   Thus   we consider  final states associated 
  with the production of  two 
  jets widely separated in rapidity~\cite{muenav,epr1012}. 
To be specific,   we consider    correlations of   
a forward and a central jet   (Fig.~\ref{fig:jetcorr}), 
and investigate  the associated  transverse energy flow   
as a function of  pseudorapidity and 
azimuthal angle in the transverse plane~\cite{epjc12}.

The region of high    rapidities is critical.  
While   first measurements of forward jet spectra 
 at the LHC~\cite{cms-prelim} are roughly in agreement 
with predictions from different  Monte Carlo   simulations,   detailed 
aspects of production rates   and  correlations~\cite{cms-prelim,dijetratios}    
are not  well    understood yet.  
From   the underlying event standpoint~\cite{atlas-ue11,cms-ue11},   
 energy flow measurements~\cite{cms-pas-10-02} 
  in minimum bias and dijet events  
emphasize the  difficulty~\cite{bartfano1103}  in achieving   a unified 
underlying event description 
from central to forward rapidities based on \pythia~\cite{pz_perugia} Monte Carlo 
tuning. 

Ref.~\cite{epjc12} considers production 
of central and forward jets   
(taking e.g. central and forward jet   pseudorapidities
in the range  $  1 <  \eta_c  <  2  , \;\;  - 5  <  \eta_f  <  - 4    $),  
and  the   transverse energy flow 
\begin{equation}
{ { d E_\perp } \over { d \eta}  }   =  { 1 \over \sigma}  \int dq_\perp \   q_\perp  \  
{ { d \sigma  } \over {dq_\perp \    d \eta}  }   \;\;   .   
\nonumber
\end{equation}
While the measurements~\cite{cms-pas-10-02} are designed to investigate 
properties of the soft underlying event,  this energy flow  observable  is 
sensitive to  harder   color radiation.  Also, it enables one to 
access   more details on the  structure  of the final states associated with  the 
jet production processes  observed  in~\cite{cms-prelim,dijetratios}. 
The  transverse  factor 
$q_\perp   $   in the above energy flow  distribution  
 enhances  matrix element corrections 
 due to     extra hard-parton emission at short distances,  and 
gives contributions which break the transverse 
momentum ordering  approximation in the long-distance evolution 
of the  parton  showers.   Ref.~\cite{epjc12}  computes these effects in the  
high-energy  factorization framework~\cite{epr1012,hef}. 

The transverse energy flow, obtained by summing  the energies over all particles 
in the final states, is naturally  also sensitive
  to soft particles being produced into the  final states.  In order to study  
  hard   radiation   one may  rather consider 
  the associated   charged particle  p$_T$ spectra.  However,  
   at the LHC it is possible to control the infrared sensitivity  of the energy flow 
by looking  at an alternative  observable, defined in a  
 different manner~\cite{epjc12} as follows.   
 One may first  cluster  particles into jets  by means of a  jet algorithm, 
 and then  construct   the associated energy flow from  jets  with transverse energy 
 above a given lower bound $q_0$.  Infrared safety is ensured by   running  
a jet algorithm, as opposed to applying the bound on the 
energy flow integral.  The question is  which value of  $q_0$ is 
phenomenologically  meaningful.  At the LHC the transverse energy per unit rapidity 
is large enough  across a wide rapidity range  that  a mini-jet type of 
bound $q_0 \approx  5$ GeV  should be    feasible.  
 This is to be contrasted with previous 
collider experiments, where one either did not have the detector capabilities to 
go very   forward  in rapidity  (as at the Tevatron)    or  did not have 
  enough  transverse energy per unit rapidity (as at HERA, about 
  $1 \div 2$ GeV  per unit rapidity).  Calorimetric  measurements of  
  this mini-jet energy  flow at the LHC  will be   interesting. 
 
\begin{figure}[t!]
\vspace{4.5cm}
  \begin{picture}(30,0)
    \put(40, -40){
      \includegraphics{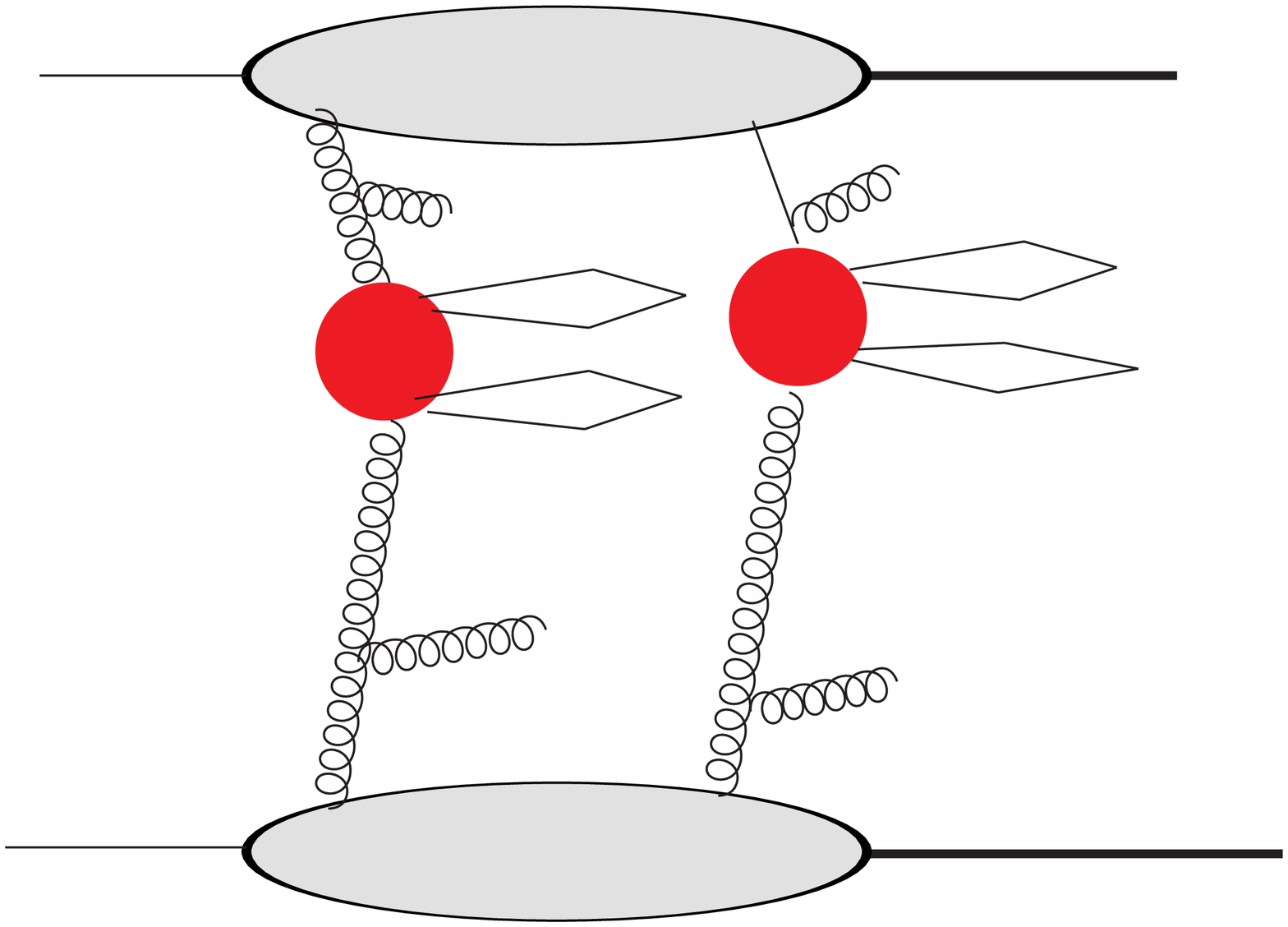}
    }
    \put(270, -40){
      \includegraphics{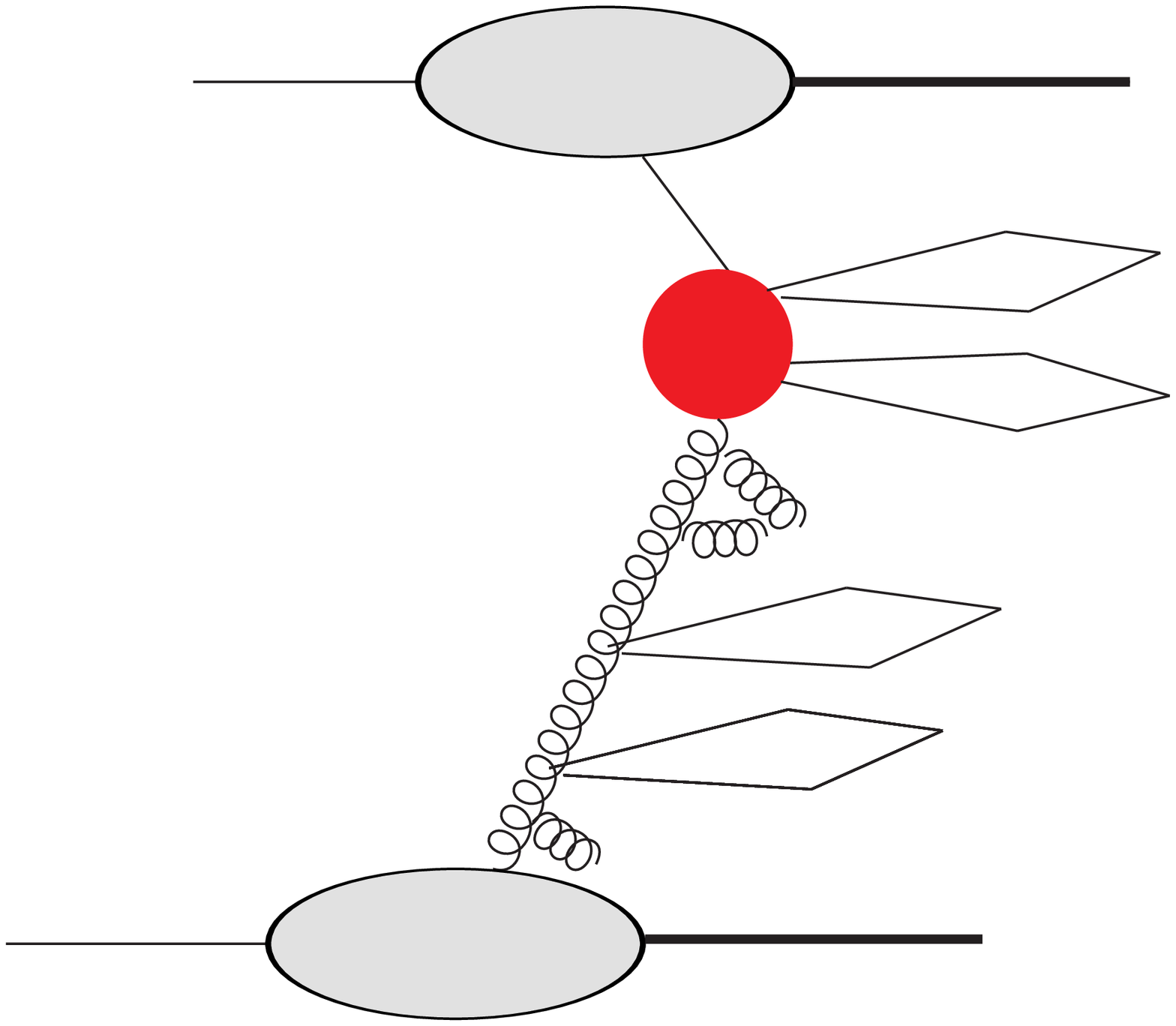}
    }    

     \end{picture}
\caption{\it Multi-jet production by 
 (left) multiple  parton collisions; (right) single parton collision.} 
\label{fig:mpi-jet}      
\end{figure}

Multiple parton collisions (Fig.~\ref{fig:mpi-jet})   form 
 one of the major    motivations  for such    energy flow studies.   
Multiple collisions   become increasingly important with energy 
 as parton densities grow~\cite{pavtrel82},  contributing primarily to  highly 
 differential cross sections sensitive to the detailed distribution of the states produced 
 by parton evolution.   Their  role at the LHC  is being studied very actively 
 both by experiment~\cite{bartal,cms-ue11,bartfano1103} and 
 theory~\cite{bartal,markus-talk,yuri-talk,blok}.    
 Since     multi-parton  interactions 
depend  on  the growth of   parton densities   and  probe 
 the detailed final-state structure, 
 their treatment   should  be   affected   
 by    corrections  to    parton  shower evolution.  
Collinear ordering   is known to give an effective 
 picture of parton evolution  for inclusive observables; however,  
 it is not expected to represent    the detailed final states reliably 
 when longitudinal momentum fractions $x$ become small and  
 parton densities increase.  So,    in particular,    noncollinear   
high-energy  corrections to   QCD showers  could   affect 
the analysis  of multiple interactions significantly~\cite{ajaltouni,hj_rec}.   
The energy flow  in  forward-central 
jet production    may provide a first step  to analyze this issue.

\begin{figure}[htb]
\vspace{60mm}
\includegraphics{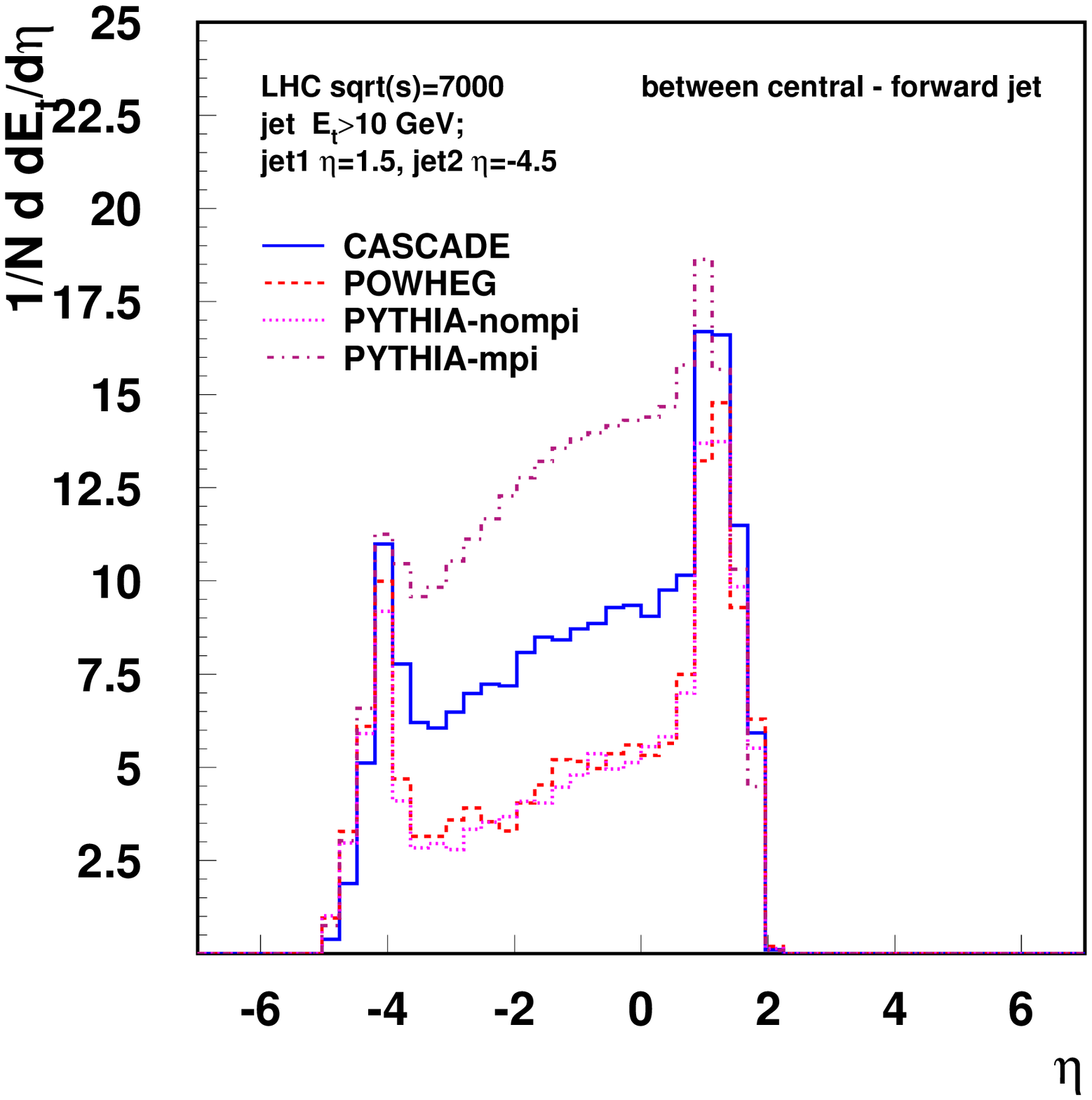}
\includegraphics{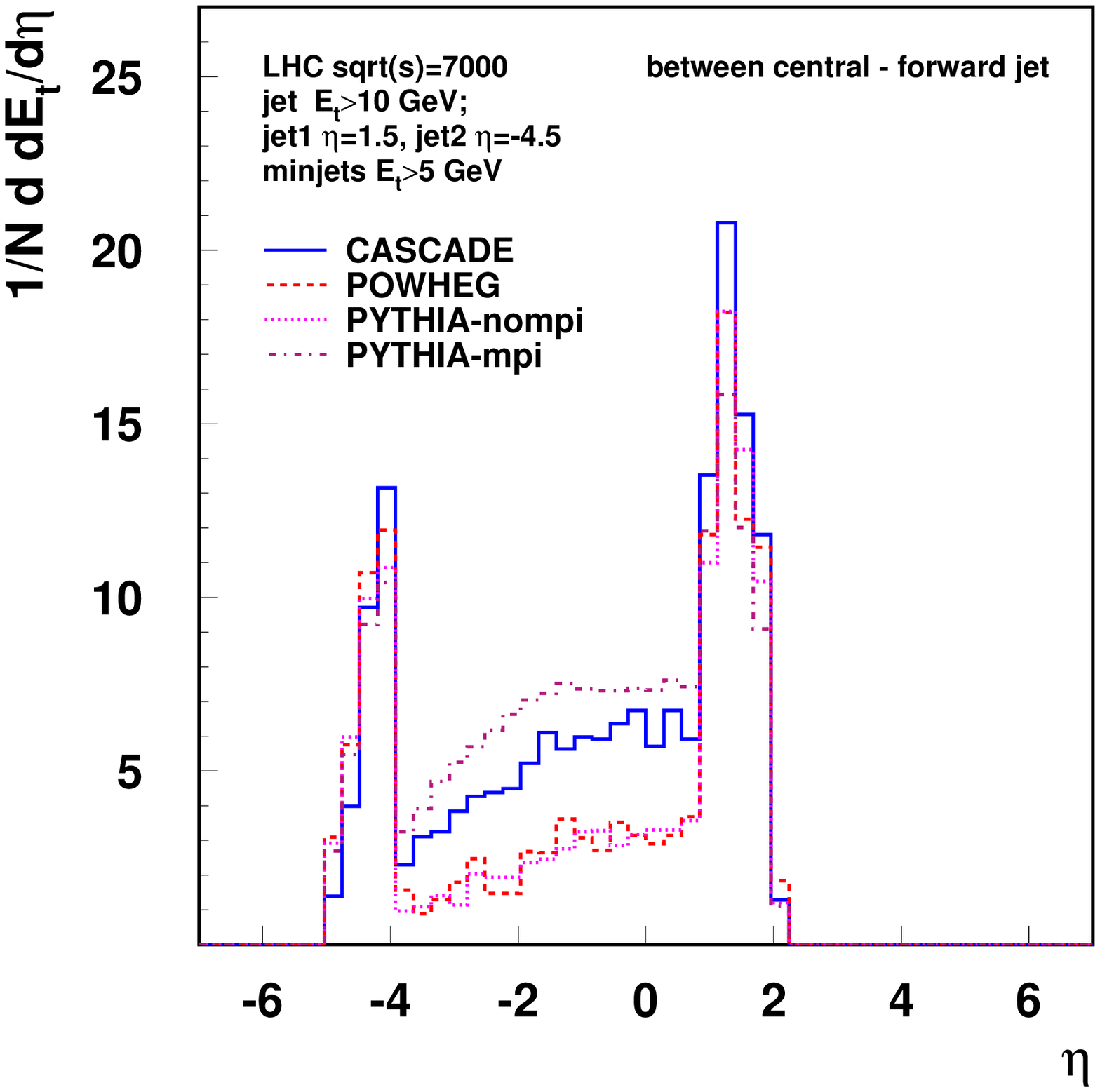}  
\caption{\it  Transverse 
energy flow~\cite{epjc12} in the  inter-jet   region:  (left) particle flow; (right) mini-jet flow. } 
\label{fig:betw} 
\end{figure}

\begin{figure}[htb]
\vspace{53mm}
\includegraphics{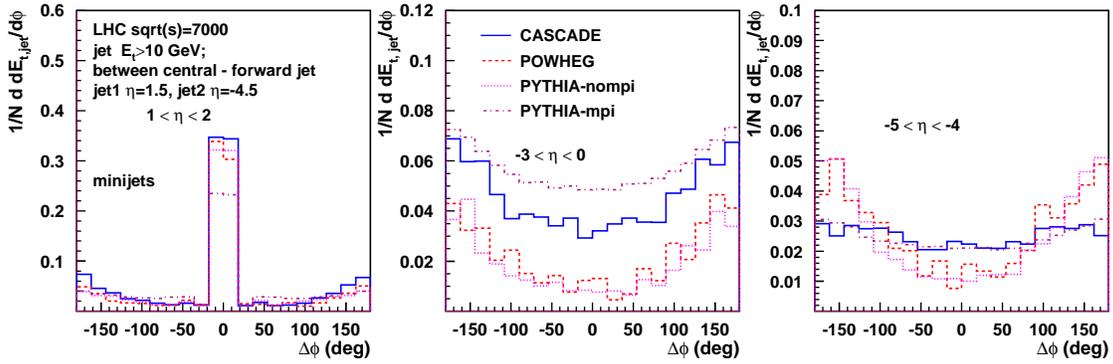}
\caption{\it  Azimuthal dependence of the  mini-jet   
energy flow~\cite{epjc12}   for different rapidity ranges:   
(left)  central-jet;   (middle) intermediate;  (right) forward-jet. 
  } 
\label{fig:azim} 
\end{figure}

Figs.~\ref{fig:betw}   and~\ref{fig:azim}
report results  for  the energy   flow~\cite{epjc12}    
 from three Monte Carlo event generators: 
 the  k$_\perp$-shower \cascade\  generator~\cite{cascade_docu},   to 
 evaluate contributions  of 
  high-energy logarithmic corrections; the NLO matched \powheg\  generator~\cite{alioli}, 
  to evaluate the  effects of  NLO   corrections to matrix elements; 
   \pythia\  Monte Carlo~\cite{pz_perugia},  
 used in two different modes:  with    the LHC tune  Z1~\cite{rickstune}  
  (\pythia-mpi) to evaluate  contributions of 
   multi-parton interactions, 
and without  any  multi-parton interactions (\pythia-nompi).  

Fig.~\ref{fig:betw}  shows the pseudorapidity dependence 
 of the  transverse energy flow  in the 
region between the central and forward jets. 
The particle energy flow   plot     on the left in Fig.~\ref{fig:betw}   
shows  the  jet profile picture,  and indicates     enhancements    of  
the energy flow   in the inter-jet region 
with respect to the \pythia-nompi    result 
 from higher order emissions in  \cascade\  and from multiple parton collisions in 
 \pythia-mpi. On the other hand, there is  little effect 
   from  the next-to-leading 
 hard correction  in \powheg\   with respect to  \pythia-nompi. 
  The energy flow  is dominated by
multiple-radiation, parton-shower effects.   The mini-jet energy flow  plot   on the right in 
Fig.~\ref{fig:betw}   
  indicates similar 
effects,   with reduced sensitivity  to infrared radiation. 
As the mini-jet flow definition suppresses the
contribution of soft radiation,    the  \cascade\  and  \pythia-mpi results become 
more similar in the inter-jet  region.  
Distinctive effects  are also found in~\cite{epjc12} by computations in the region away 
from the jets.

  Fig.~\ref{fig:azim}    illustrates  the azimuthal dependence 
of the mini-jet transverse energy flow. 
Here $\Delta \phi$   is  measured  with respect to the central jet.   The 
$\Delta \phi$  distribution is shown for three different rapidity ranges, 
corresponding  to the central-jet,  forward-jet,    and intermediate  rapidities. 
As we go  toward  forward rapidity,   the 
 \cascade\  and  \pythia-mpi calculations  give   a more   
 pronounced flattening  of the $\Delta \phi$  distribution compared 
 to  \powheg\   and   \pythia-nompi,    
  corresponding to   increased decorrelation between the jets.

The above  numerical results  indicate that  quite  distinctive 
behaviors   should  be expected from  
measurements of  particle and mini-jet energy  
flows associated with  production of forward and central jets. 
They  will  tell us  about several soft-physics effects,  
 from  the structure of underlying events  to  
multiple parton collisions to QCD showering, 
which are relevant to a range of subjects in LHC physics: 
from     studies  of  color flow  
in  the QCD tuning of Monte Carlo event 
generators  to searches  for  new physics signals  based on  
the structure   of   jets.  
One feature emerging already from the above studies is that  
gluon emission over large rapidity intervals  gives  sizeable 
contribution to the inter-jet  energy flow. As a result, 
the rates for    multi-parton interactions 
may  be influenced significantly  by non-collinear corrections  to single-chain 
showering.   
From the theory viewpoint, it  underlines the relevance of approaches 
which aim at  a  more accurate and complete   description of  
 initial state dynamics  by generalizing the 
notion of parton distributions, both 
for  quark-dominated~\cite{becher-neub} and gluon-dominated~\cite{xiao} processes.

\vskip 0.8 cm

\noindent 
{\bf Acknowledgments}.  
I thank the convenors   for the   
 invitation  to a very nice  workshop. 
The results in this article have been obtained in 
collaboration with    M.~Deak,       H.~Jung and K.~Kutak.



\begin{footnotesize}

\end{footnotesize}

\end{document}